\newcommand{\be}{\begin{equation}}
\newcommand{\bea}{\begin{eqnarray}}
\newcommand{\ee}{\end{equation}}
\newcommand{\eea}{\end{eqnarray}}
\newcommand{\nn}{\nonumber}

\newcommand{\qa}{\alpha}
\newcommand{\qb}{\beta}
\newcommand{\qg}{\gamma}

\newcommand{\qd}{\delta}
\newcommand{\qD}{\Delta}
\newcommand{\qe}{\varepsilon}

\newcommand{\qz}{\zeta}

\newcommand{\qy}{\theta}

\newcommand{\qk}{\kappa}
\newcommand{\ql}{\lambda}

\newcommand{\qr}{\rho}
\newcommand{\qs}{\sigma}

\newcommand{\qt}{\tau}
\newcommand{\qf}{\varphi}
\newcommand{\qF}{\Phi}

\newcommand{\qj}{\psi}

\newcommand{\qo}{\omega}

\newcommand{\tr}{{\rm tr}}

\newcommand{\rd}{{\rm d}}

\newcommand{\EE}{{\mathbb E}}

\newcommand{\NN}{{\mathbb N}}

\newcommand{\RR}{{\mathbb R}}

\newcommand{\sH}{{\sf H}}
\newcommand{\bits}{ \{0,1\} }
\newcommand{\cA}{{\mathcal A}}

\newcommand{\cK}{{\mathcal K}}

\newcommand{\cN}{{\mathcal N}}
\newcommand{\cO}{{\mathcal O}}

\newcommand{\cR}{{\mathcal R}}

\newcommand{\cX}{{\mathcal X}}
\newcommand{\cY}{{\mathcal Y}}

\newcommand{\vecm}{{\boldsymbol{m}}}

\newcommand{\vecN}{{\boldsymbol{N}}}

\newcommand{\vecx}{{\boldsymbol{x}}}
\newcommand{\vecX}{{\boldsymbol{X}}}

\newcommand{\vecy}{{\boldsymbol{y}}}
\newcommand{\vecY}{{\boldsymbol{Y}}}
\newcommand{\vecz}{{\boldsymbol{z}}}
\newcommand{\vecZ}{{\boldsymbol{Z}}}
\newcommand{\vecnu}{{\boldsymbol{\nu}}}

\newcommand{\vecqz}{{\boldsymbol{\qz}}}

\newcommand{\pr}{{\rm Pr}}

\newcommand{\acc}{{\rm acc}}
\newcommand{\Nacc}{N_{\rm acc}}

\newcommand{\ketbra}[1]{| #1 \rangle\langle #1 |}

\newcommand{\isdef}{\stackrel{\rm def}{=}}

\documentclass[10pt]{article}

\usepackage{graphics}
\usepackage{graphicx}
\usepackage{url}
\usepackage{amssymb}
\usepackage{amsmath}
\usepackage{a4wide}
\usepackage{enumitem}
\usepackage{xcolor}
\usepackage{booktabs}
\usepackage{multirow}
\usepackage[colorlinks=true, citecolor=blue, linkcolor=blue, urlcolor=blue]{hyperref}
\usepackage{array}

\begin{document}

\title{Random coding for long-range Continuous-Variable QKD}

\author{Arpan Akash Ray, Boris \v{S}kori\'{c}\\
Eindhoven University of Technology}

\date{ }

\maketitle

\begin{abstract}
\noindent
Quantum Key Distribution (QKD) schemes are key exchange protocols 
based on the physical properties of quantum channels.
They avoid the computational-hardness assumptions that underlie the security of classical key exchange.
Continuous-Variable QKD (CVQKD), in contrast to qubit-based discrete-variable (DV) schemes, makes use of quadrature measurements of the electromagnetic field.
CVQKD has the advantage of being compatible with standard telecom equipment, but at long distances
has to deal with very low signal to noise ratios, which necessitates labour-intensive error correction.
It is challenging to implement the error correction decoding in realtime.

In this paper we introduce a random-codebook error correction method that is suitable for long range Gaussian-modulated CVQKD.
We use likelihood ratio scoring with block rejection based on thresholding.
For proof-technical reasons, the accept/reject decisions are communicated in encrypted form;
in this way we avoid having to deal with non-Gaussian states in the analysis of the leakage.
The error correction method is highly parallelisable, which is advantageous for realtime implementation.
Under conservative assumptions on the computational resources, we predict a realtime key ratio  
of at least 8\% of the Devetak-Winter value,
which outperforms existing reconciliation schemes.
\end{abstract}

%========================================================================
%========================================================================

\setlength{\parindent}{0mm}

\newtheorem{theorem}{Theorem}[section]
\newtheorem{corollary}[theorem]{Corollary}
\newtheorem{lemma}[theorem]{Lemma}
\newtheorem{definition}[theorem]{Definition}
\newtheorem{proposition}[theorem]{Proposition}

%\counterwithin{figure}{section}
%\setcounter{figure}{0}
%\renewcommand{\thefigure}{\arabic{section}.\arabic{figure}}
%\renewcommand{\thetable}{\arabic{section}.\arabic{table}}

%===========================================
\section{Introduction}

%---------------------------------------------------
\subsection{Error correction in Continuous-Variable QKD}

Quantum Key Distribution (QKD) is a family of protocols that 
leverages the inbuilt tamper-evidence of quantum channels to 
generate a shared cryptographic key between two parties.
The main appeal of QKD is that its security does not depend on the type of computational-hardness assumptions 
that underlie key exchange protocols like the NIST standard Crystals-Kyber KEM
\cite{CrystalsKyber}.
The computational problems on which post-quantum key exchange schemes are based
(typically lattice problems) have not yet undergone the decades long scrutiny 
necessary to build confidence. 
Drawbacks of QKD are limited range, the need for special equipment, the need to defend against optical side channel attacks, and immaturity of the technology and certification infrastructure.
For background information on 
QKD use case rationales and recommendations we refer to 
\cite{QKDuse,RennerWolf2023}.

Continuous-Variable Quantum Key Distribution (CVQKD)
is a variant of QKD that works with weak coherent states and quadrature measurements.
It is an attractive QKD variant because of its cost-effectiveness, which
is due to the use of standard telecom equipment.
In particular, discrete-variable variants face the problem that
single-photon detectors are expensive.

In Discrete-Variable QKD (DVQKD), propagation of light over long distances causes
photon loss, which leads to non-detection events at the receiver side.
The loss is very easy to deal with: the event is simply discarded.
In CVQKD, in contrast, a quadrature observation always results in a measurement outcome.
The loss of photons causes a deterioration of the signal-to-noise ratio (SNR) at the receiver's side.
The extremely low SNR has to be dealt with by using high-performance error correcting codes.
Various codes have been considered for CVQKD.
Popular types of code are LDPC codes \cite{JCL2016,MFZG2017,JYHZ2018,Guo2020,Gumus2021}, 
Turbo codes \cite{Lodewyck2007AllFiber,VanAssche2004GaussianReconciliation}
and Polar codes \cite{JK2014,ZHFJ2021},
all of which approach the Shannon limit.
(See \cite{YYYLLCML2023} for an overview.)
An important development in CVQKD was the introduction of \emph{multidimensional reconciliation} \cite{LABZG2008} 
 in which Gaussian variables are mapped into higher-dimensional hypercubic constellations that closely approximate binary-input AWGN channels.

Table~\ref{tab:cvqkd_ec_long_distance} shows an overview of reported error correction systems and the claimed key ratios
at various distances. 
`DW' stands for the Devetak-Winter value of the key ratio (see Section~\ref{sec:prelim}),
and \%DW is the attained percentage of the DW value. 

One of the main parameters in error correction is the reconciliation efficiency, often denoted as $\qb$, which denotes 
the code rate compared to the channel capacity.
For turbo codes, efficiencies  $\beta \approx 70$–$80\%$ have been reported \cite{Lodewyck2007AllFiber,VanAssche2004GaussianReconciliation}
in QKD.
Low-rate LDPC codes routinely achieve $\beta \approx 95$–$98\%$,
polar codes $\beta \approx 94$–$96\%$,
and the rateless Raptor code up to $\beta \approx 98\%$.
The discrepancy between the high $\qb$ and low \%DW is caused by the rejection of blocks (often called {\em frame errors}). 
We note that discarding frames is a form of post-selection \cite{postselection2013}, which affects Eve's state and hence
influences the key leakage in a complicated way.
Post-selection has to be accounted for carefully in the security proof. 
As far as we can discern this has not been done in the works cited in Table~\ref{tab:cvqkd_ec_long_distance},
and hence it is not clear what their actual secret key ratios are (other than \cite{hajomer100km}, which has zero frame error rate).

In QKD it is important not only to be able to correct the errors, but also to do it {\em in real time}.
This is a particularly tough challenge for the decoding step, which is the most computationally demanding.
LDPC decoding, for instance, at low rate needs many iterations which cannot be done in parallel.
As can be seen in Table~\ref{tab:cvqkd_ec_long_distance},
works on long-distance CVQKD either do not try to implement realtime decoding 
or manage to realize it only for relatively short distance.
Our aim is to develop a reconciliation scheme for long-distance CVQKD 
that allows for realtime decoding.

\begin{table}[t]
\caption{\textit{
CVQKD at various distances. 
$T\in(0,1)$ is the attenuation parameter of the quantum channel.
The losses marked with `$*$' or `$**$' come from an ultra low-loss optical fiber, with loss rated at $0.16$ dB/km and $0.14$ dB/km respectively (for $1550$ nm). 
The rest use a more standard optical fiber at loss $0.2$ dB/km.}
}
\centering
\small
\begin{tabular}{|>{\centering\arraybackslash}p{1.4cm}|>{\centering\arraybackslash}p{1.45cm}|>{\centering\arraybackslash}p{1.1cm}|>{\centering\arraybackslash}p{1.1cm}|>{\centering\arraybackslash}p{1.0cm}|>{\centering\arraybackslash}p{1.4cm}|>{\centering\arraybackslash}p{1.4cm}|>{\centering\arraybackslash}p{0.8cm}|>{\centering\arraybackslash}p{0.9cm}|}
\toprule

\multirow{2}{*}{Paper}&\multirow2{*}{Code}&
Distance (km)  &
\multirow{2}{*}{$T$} &
Loss (dB)&
Key bits/s &
\multirow{2}{*}{Pulses/s} &
\multirow{2}{*}{\%DW} & Real-time? \\
\midrule
% --- Polar (first)
\cite{Zhang2020PRL202km}&Polar  & $27$ & $0.37$    & $4.4$*  & $2.8\times 10^{5}$ & $5.0\times 10^{6}$ & $17$  & No \\
\cite{Zhang2020PRL202km}&Polar  & $49$ & $0.15$    & $8.3$* & $6.2\times 10^{4}$ & $5.0\times 10^{6}$ & $11$  & No \\
\midrule
% --- LDPC (second)
\cite{frazao2024aqis} &LDPC   & $13$  & $0.56$   & $2.6$   & $5.8\times10^6$   & $7.5\times 10^{7}$ & $13$ & No\\
\cite{Jouguet2013NatPhotonics80km} &LDPC   & $25$  & $0.32$   & $5.0$   & $10^{4}$   & $1.0\times 10^{6}$ & $3.7$ & Yes\\
\cite{frazao2024aqis} &LDPC   & $26$  & $0.31$   & $5.1$   & $5\times10^5$   & $7.6\times 10^{7}$ & $2.5$ & No\\
\cite{Wang2015HighRate}&LDPC   & $50$ & $0.10$   & $10$ & $5.2\times 10^{4}$ & $2.5\times 10^{7}$ & $2.7$ & No \\
\cite{Zhang2020PRL202km}&LDPC   & $70$ & $0.068$  & $12$*  & $4.3\times 10^{4}$ & $5.0\times 10^{6}$ & $17$ & No \\
\cite{Jouguet2013NatPhotonics80km}&LDPC   & $81$ & $0.025$  & $16$  & $5\times 10^{2}$ & $1.0\times 10^{6}$ & $2.8$ &Yes \\
\cite{hajomer100km}&LDPC   & $100$ & $0.029$  & $15$**  & $10^{5}$ & $1.0\times 10^{8}$ & $4.7$ & No \\
\cite{Zhang2020PRL202km}&LDPC   & $141$ & $0.0045$ & $23$*  & $3.2\times 10^{2}$ & $5.0\times 10^{6}$ & $2.0$ & No \\
\midrule
\cite{Zhang2020PRL202km}&Raptor & $203$ & $0.00057$ & $32$*  & $6.2$ & $5.0\times 10^{6}$ & $0.30$ & No  \\
\midrule
\multirow{3}{*}{Our work}& random codebook ($q=2^{15}$)  & \multirow{3}{*}{$300$} & \multirow{3}{*}{$10^{-6}$} & \multirow{3}{*}{$60$}  & \multirow{3}{*}{$0.059$} & \multirow{3}{*}{$1.0\times 10^{6}$} & \multirow{3}{*}{$8.2$} & \multirow{3}{*}{Yes} \\
\bottomrule
\end{tabular}
\label{tab:cvqkd_ec_long_distance}
\end{table}

%-------------------------------------------------------------------------
\subsection{Contributions and outline}

We propose an information reconciliation scheme 
for long range Gaussian-modulated CVQKD, based on random codebooks.
In contrast to most of the literature, we take into account
that the existence of a frame rejection mechanism 
conditions the relevant part of Eve's state on the `accept' decision,
which affects the leakage analysis and hence the provably secure key rate.

We first present a simple-to-analyze but impractical variant.
In a nutshell, the steps are as follows.
Alice sends multiple blocks of $n$ Gaussian-sampled coherent states; 
the sequence of quadratures in a block is denoted as~$X$.
For each block, Bob's measurement outcome is a sequence $Y$ of quadratures.
Bob hides each $Y$ in a big $q$-element table that contains random sequences,
drawn from the same distribution of $Y$, i.e. the fake entries are indistinguishable from 
$Y$ for a party who has no side information about~$X$.
Bob sends all the tables to Alice.
Alice needs to find out which entry in each table is the `real' one;
this represents $\log_2 q$ bits of information per block.
Alice tests each table entry, using a similarity score between $X$ and the entry, essentially using MAP decoding
with a fixed threshold.
If a {\em single} score exceeds the threshold, the block is accepted and the index of the winning entry
serves as the error-corrected $q$-ary symbol; if zero or multiple scores exceed the threshold, the block is rejected.
Next Alice tells Bob which blocks were accepted.
This communication is one-time-pad (OTP) encrypted, and consumes pre-existing shared key material.
The accepted blocks may still contain a few errors due to blocks that were wrongly decoded by Alice.
In order to get rid of these errors, Alice and Bob apply an ordinary error-correcting code,
e.g.~a binary linear code or a Reed-Solomon code.
This is done in a way that hides the number of accepted blocks.

The encryption of Alice's accept/reject decisions is a crucial point and
needs to be carefully motivated.
The reason why we want to perform this encryption is proof-technical:
{\em If the decisions would be sent in the clear, the state of Eve's quantum memory would get conditioned on the fact of acceptance.}
The conditioned state would no longer be Gaussian but a mixture of Gaussian states,
making the security analysis intractable.
Hiding all the decisions from Eve reduces the analysis of the leakage to well established equations
that involve only Gaussian states.
But there is an important caveat here: 
it is not at all trivial to hide the {\em number of accepted blocks} $N_\acc$ from Eve,
since she may learn this piece of information by other means, e.g.~by observing how much key material Alice and Bob are
using after the QKD.
Therefore we see our protocol in the context of a hybrid key exchange that combines key material from 
QKD and a post-quantum KEM, in such a way that a final key is derived whose length does not depend on the
amount of key material from the QKD.
In this context it is safe to assume that Eve has no way to learn $N_\acc$ from a side channel.

We study the {\em secret key ratio} (SKR) of the proposed scheme, which is defined as
the QKD key length divided by the number of laser pulses.\footnote{
For the purpose of analyzing the reconciliation performance, 
the key ratio is a more useful metric than the key rate (key bits per second), 
as it is insensitive to equipment parameters such as number of laser pulses per second. 
}
In the calculation of the key ratio, we have to subtract the OTP key expenditure from the QKD key length.
We work in the regime of large number of blocks, so that the reconciliation efficiency of the ordinary code
approaches the asymptotic value.
We obtain SKR values that are several percent of the Devetak-Winter value.
The decoding procedure is highly parallelisable, which helps to achieve a realtime implementation.

\vskip1mm

We introduce a modified scheme version that has better communication complexity.
Instead of sending entire tables, Bob sends seeds from which pseudorandom tables are
generated.
The resulting protocol resembles reverse reconciliation systems that have
Bob xor-ing a randomly chosen codeword from a {\em structured} ECC into his measurement sequence~$y$.
We discuss options for efficient parallel implementation of the decoding.

The organisation of the paper is as follows.
In Section~\ref{sec:prelim} we introduce notation and
present some background on CVQKD with Guassian modulation.
In Section~\ref{sec:protocol} we present the `impractical' version of our protocol,
with truly random tables.
Section~\ref{sec:stat} contains the statistical analysis of the scheme,
including the acceptance probability and the symbol error rate within the set of accepted symbols.
In Section~\ref{sec:SKR} we present the SKR results.
In Section~\ref{sec:imp} we introduce the more practical pseudorandomness-based version of the scheme
and discuss efficient implementation options.
Section~\ref{sec:discuss} concludes with a discussion and topics for future work.

%===========================================
\section{Preliminaries}
\label{sec:prelim}

%-------------------------------------------------------------------------
\subsection{Notation}

For integer $q$, we write $[q]$ as shorthand for the set $\{1,\ldots,q\}$.
Stochastic variables are written in capital letters, 
and their numerical values in lowercase.
The notation $\pr[X=x]$ stands for the probability that the random variable $X$ 
takes value~$x$.
The expectation over $X$ is defined as $\EE_x f(x)=\sum_x \pr[X=x]f(x)$.
The normal distribution is denoted as  $\qf(x)=\frac1{\sqrt{2\pi}}e^{-x^2/2}$.
Its cumulative distribution is written as 
$\qF(x) = \int_{-\infty}^x\rd u\; \qf(u)$.

The Shannon entropy of a random variable $X$ is denoted as $\sH(X)$.
We write $h$ for the binary entropy function, $h(p)=p\log_2\frac1p+(1-p)\log_2\frac1{1-p}$.
`ln' is the natural logarithm. 
We measure all entropies in bits.
The mutual information between $X$ and $Y$ is written as $I(X;Y)$.
The entropy of $Y$ given $X$ is written as $\sH(Y|X)=\EE_x\sH(Y|X=x)$.
A quantum state is described by a density matrix~$\qr$. 
We write $S(\qr)=-\tr\qr\log_2\qr$ for the von Neumann entropy of~$\qr$.
Consider a joint classical-quantum state of a classical random variable $Y\in\cY$
and a quantum system E that is correlated with~$Y$.
The marginal state of $Y$ is written as a quantum state 
$\qr^Y=\sum_{y\in\cY}\pr[Y=y]\ketbra y$ representing the classical distribution.
For the uniform distribution we use the notation $\qt=\sum_y \frac1{|\cY|}\ketbra y$.
The joint state is written as $\qr^{YE}=\sum_{y\in\cY}\pr[Y=y] \ketbra y\otimes \qr^E_y$,
where $\qr^E_y$ is the state of the E system conditioned on~$Y=y$.
The von Neumann entropy of the quantum system given $Y$ is denoted as 
$S(E|Y)_\qr=\EE_y S(E|Y=y)_\qr = \EE_y S(\qr^E_y)$.
The mutual information between E and $Y$ is
$I(E;Y)_\qr=S(E)_\qr-S(E|Y)_\qr$.

As the figure of merit of a CVQKD scheme we use the Secret Key {\bf Ratio}
(SKR), which equals the length of the constructed secret key
divided by the number of laser pulses.
This is different from the key {\em rate} (key bits per second).
With the key {\em ratio} it is possible to do a fair comparison between 
post-processing schemes of CVQKD setups
that have a different number of pulses per second.
Note that we do not take into account any pulses that are used solely for
channel monitoring.

%-------------------------------------------------------------------------
\subsection{CVQKD with Gaussian modulation, homodyne detection, and reverse reconciliation}

We briefly review CVQKD basics.
We consider Gaussian modulation, for simplicity.
We consider the usual attacker model: Eve can manipulate quantum states in transit,
but cannot access Alice and Bob's laboratories.
Alice and Bob communicate over a classical channel that is
noiseless, public, and authenticated. 

Alice draws Gaussian $P, Q\in\RR$ with zero mean and variance~$\qs_X^2$.
She sends a coherent state with quadrature values centered on $(P,Q)$.
Bob randomly measures one of the two quadratures.
This procedure is repeated $n$ times.
Bob's outcomes are $Y\in\RR^n$.
Alice's displacement components that match Bob's choice of quadratures 
are denoted as $X\in\RR^n$.

Alice and Bob monitor the noise in the quantum channel.
The channel can be characterised by two parameters \cite{leverrier:tel-00451021}:
the attenuation $T$ and the excess noise power~$\xi$.
It holds that $y_i = \sqrt T x_i+N_{\rm shot}+N_{\rm exc}$ where 
$N_{\rm shot}$ is shot noise, Gaussian  with zero mean and variance $\frac12$;
the $N_{\rm exc}$ is excess noise, Gaussian with zero mean and variance~$T\xi/2$.
We do not include the quantum efficiency and electronics noise
of the detector in the model, as they can be absorbed in~$T$  \cite{YMKSK2023}
due to the noise-loss equivalence.
This remodeling leads to a small value of the remaining excess noise power $\xi$;
in the remainder of the paper we will set $\xi$ to be negligible.
The variance of $Y_i$ is given by
\be
    \qs_Y^2 = T\qs_X^2 +\frac12+\frac12 T\xi
    \quad\quad\quad
    \qs_{Y|X}^2 = \frac12+\frac12 T\xi.
\ee
In long-distance CVQKD we have $T\ll 1$.
We consider the regime where $T\qs_X^2< 1$, $\xi \ll \qs_X^2 $, and we write
the signal to noise ratio as
\be
    \qe\isdef \frac{T\qs_X^2}{\qs_{Y|X}^2}
\ee
with $\qe\ll 1$.
Bob informs Alice of his quadrature choices.
Bob generates a secret key $K\in\cK$, which typically is a hash of some $U$, where $U$ is
an error-correction codeword or a discretisation of~$Y$.
He computes information reconciliation data $P$ and sends $P$ to Alice.
Typically $P$ contains a seed for the hash and e.g.\;the syndrome of the discretised $Y$,
or a codeword xor'ed into~$Y$.
From $X$ and $P$ Alice reconstructs~$K$ by first reconstructing $U$ and then hashing~$U$ with the provided seed.

It has been shown that the optimal attack against Gaussian-modulated CVQKD is a Gaussian attack,
i.e.\;resulting in a Gaussian state.
Furthermore, it suffices to consider collective (i.i.d.) attacks \cite{gaussianoptimal, CVdeFinettiLev2017}.
The maximum achievable size of $K$, in bits, divided by $n$ is called the secret key capacity.
It has been shown \cite{DW2005} that the secret key capacity for Gaussian modulation and one-way reverse reconciliation
($C_{\rm secr}^{\rm Gauss}$)
is bounded as
\be
    C_{\rm secr}^{\rm Gauss} \geq \triangle I\isdef I(X_i;Y_i) - I(E_i;Y_i).
\label{Capacity}
\ee
The mutual information between the classical $X_i$ and $Y_i$ follows from the signal-to-noise ratio,
\be
    I(X_i;Y_i) = \frac12 \log_2(1+\qe )
    = \frac\qe{2 \ln 2}- \cO(\qe^2).
\label{IXiYi}
\ee
The leakage $I(Y_i;E_i)=S(E_i)-S(E_i|Y_i)$  is discussed in Appendix~\ref{app:leak}.
For low SNR it is approximately given by
\be
    I(E_i;Y_i) =  \frac\qe{2\ln 2}  \cdot\qs_X^2 \ln \frac{1+\qs_X^2}{\qs_X^2}
    +\cO(T\xi \log\frac1{T\xi})- \cO(\qe^2). 
\label{IEiYi}
\ee
In the limit of large modulation variance it holds that
\be
	I(X_i;Y_i) - I(E_i;Y_i) \stackrel{\qs_X\to\infty}{\longrightarrow} \frac T{2\ln 2}+ \cO(T^2).
\label{DWvalue}
\ee 
We call $\frac T{2\ln2}$ the {\em low-T Devetak-Winter value} \cite{DW2005}. 
It is $\frac12$ of the PLOB bound \cite{PLOB2017} which represents the upper bound on the QKD key ratio.

\begin{figure}[h]
\begin{center}
\begin{picture}(200,85)
\setlength{\unitlength}{1mm}
\put(-35,-5){\includegraphics[width=.35\textwidth]{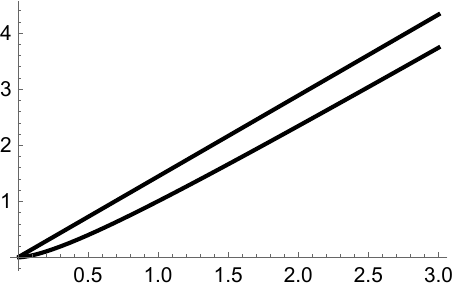}}
\put(45,-5){\includegraphics[width=.35\textwidth]{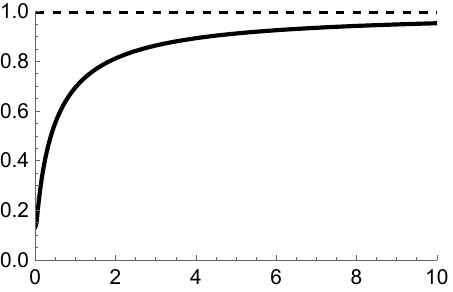}}
\put(18,-1){$\qs_X^2$}
\put(-15,12){\rotatebox{30}  {$I(X_i;Y_i)/T$}}
\put(-10,4){\rotatebox{30}  {$I(E_i;Y_i)/T$}}
\put(98,-1){$\qs_X^2$}
\put(68,19){$\frac{I(E_i;Y_i)}{I(X_i;Y_i)}$}
\end{picture}
\end{center}
%\vskip4mm
\caption{
\it 
The mutual informations $I(X_i;Y_i)$ (\ref{IXiYi}) and $I(E_i;Y_i)$ (\ref{leakFull}) as a function of the modulation variance $\qs_X^2$,
for $T\ll 1$, $\xi\ll \qs_X^2$.
}
\label{fig:II}
\end{figure}
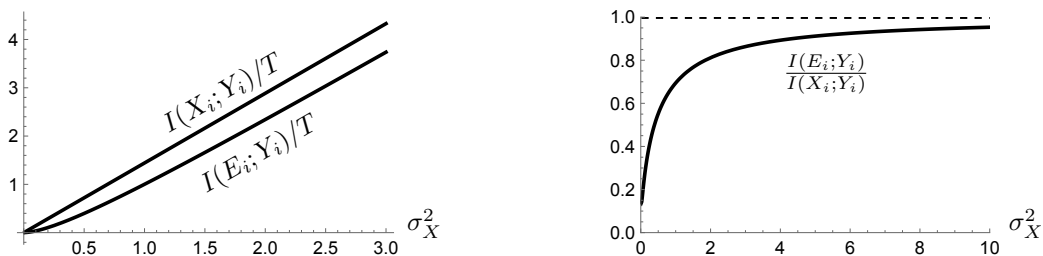

Fig.\,\ref{fig:II} shows how large the leak $I(E_i;Y_i)$ is relative to the
maximally available information $I(X_i;Y_i)$.
The gap  $I(X_i;Y_i) - I(E_i;Y_i)$ increases as a function of $\qs_X^2$ and saturates to the Devetak-Winter value.
However, in practice it is difficult to operate at large $\qs_X^2$ for two reasons:
(i)
Large modulation causes phase noise \cite{MA2017}, which manifests as excess noise;
(ii)
Working with large $\qs_X^2$ demands very high reconciliation efficiency. 
In order to have a positive key rate, the reconciliation efficiency has to be larger than
$\frac{I(E_i;Y_i)}{I(X_i;Y_i)}$.
In Fig.\,\ref{fig:II} we see that e.g.\;at $\qs_X^2\geq 9.37$ the efficiency needs to be better than~95\%.
At $\qs_X^2<1$ it is possible to obtain a positive key rate even with a highly inefficient error correction code.

%---------------------------------------------------
\subsection{`Decoupled' reconciliation data}
\label{sec:prelimdecoupled}

The SKR of CVQKD in practice is affected by 
the efficiency $\qb$ of the employed error-correction code ($\qb<1$)
and by the leakage caused by the public reconciliation data~$P$.
A high SKR is obtained
by using a highly efficient error-correcting code and leakage-free reconciliation data.
Good codes for CVQKD achieve $\qb\approx 0.95$ to~$0.98$.
Leakage-free reconciliation data is achieved by sending a one-time-pad-encrypted syndrome,
and by the trick described below.

\begin{lemma}[Lemma~1 in \cite{LABZG2008}]
\label{lemma:decoupling}
Let $U$ and $P$ be independent classical random variables.
Let $E$ be a quantum system. Then
$I(U;PE)\leq I(UP;E)$.
\end{lemma}

\begin{corollary}[Adapted from \cite{LABZG2008} to the case of reverse reconciliation]
\label{corol:decoupling}
Let $X$ and $Y$ be the measurement result of Alice and Bob respectively.
Let $E$ be Eve's quantum system.
Let $U$ be a secret chosen by Bob, independent of~$Y$. 
Let $P$ be public information reconciliation data computed from $Y$ and~$U$,
such that Alice is able to reconstruct $U$ from $P$ and~$X$. 
Let $P$ be independent from~$U$.
Then 
\be
    I(U;PE) \leq I(Y;E).
\label{Idecoupled}
\ee
\end{corollary}
\underline{\it Proof:}
Lemma~\ref{lemma:decoupling} gives $I(U;PE)\leq I(UP;E)$.
Since $P$ is computed from $U$ and~$Y$, it holds that $\sH(UP)\leq \sH(UY)$
and $I(UP;E)\leq I(UY;E)$.
Finally, we use that Bob chooses the $U$ independently of~$Y$, which allows us to write
$I(UY;E) = I(Y;E)$. 
\hfill$\square$

\vskip2mm

For the SKR in case of collective attacks this yields
\be
    R_{\rm key}^{\rm (decoupled)} \geq \qb I(X_i;Y_i) - I(E_i;Y_i) .
\label{prelimratedecoupled}
\ee
In \cite{LABZG2008} the decoupling of $P$ from $U$ is achieved by 
taking 8-dimensional subspaces and using the algebraic properties of octonions.

%-------------------------------------------------------------------------
\subsection{Useful lemmas}

\begin{lemma}[Neyman-Pearson]
\label{lemma:NP}
Let $X\in\cX$ be a random variable distributed according to $X\sim P_\qy$,
where either $\qy=0$ or $\qy=1$.
A hypothesis test function is a function $f:\cX\to \bits$ that tries to determine 
the parameter $\qy$ from the observed data~$x$.
Let $\EE_{X\sim P_0}f(X)$ be the type~I error probability of the test function, and 
$\EE_{X\sim P_1}[1-f(X)]$ the type~II error probability.
The likelihood ratio test with threshold $k$ is given by
\be
    \qf_k(x) = \left\{ \begin{array}{ll} 
    1: & \frac{P_1(x)}{P_0(x)}>k  \\
    0: & \frac{P_1(x)}{P_0(x)}<k
    \end{array}\right. .
\label{likelihoodratio}
\ee
At a fixed type~I error probability, the likelihood ratio test (\ref{likelihoodratio})
has the lowest type~II error probability of all test functions.
\end{lemma}

\begin{lemma}
\label{lemma:NPfun}
A test function obtained by applying a monotone 
%increasing 
function to the likelihood ratio 
$\frac{P_1(x)}{P_0(x)}$ and to the threshold
is equivalent to the likelihood ratio test.
\end{lemma}

%===========================================
\section{The proposed information reconciliation method}
\label{sec:protocol}

%---------------------------------------------------------
\subsection{Hiding accept/reject decisions}

As has often been noted (see e.g.\;\cite{leverrier:tel-00451021})
it is difficult to give a security proof for CVQKD schemes whose reconciliation
involves `postselection', i.e.\;rejection of blocks.
Eve's state, when conditioned on the fact that a block has been accepted,
is no longer Gaussian; instead it becomes for instance a classical mixture of Gaussian states,
which is intractable.

For a scheme with block acceptance probability $P_\acc$ it is tempting to postulate a rate formula
$R=P_\acc[\qb I(X_i;Y_i) - I(Y_i; E_i)]$ as a modification of (\ref{prelimratedecoupled}).
However, this is only allowed if the set of {\em all public data} is statistically independent of the secret key.
The accept/reject decisions for blocks do {\em not} have this independence.
For this reason we introduce a reconciliation scheme in which Alice one-time-pad encrypts all the accept/reject decisions.
This consumes pre-shared key material and hence
inflicts a heavy penalty on the achievable key ratio.

%---------------------------------------------------------
\subsection{Protocol steps}
\label{sec:steps}

\underline{Initialisation}.
Alice and Bob already have pre-shared key material.
Alice and Bob determine the channel properties $T$ and $\xi$.
They agree on the modulation variance $\qs_X^2$.
They agree on parameters $n, N \in\NN$. 
They agree on a function ${\rm Bin}$: $[q]\to \bits^{\lceil\log q\rceil}$ which maps $q$-ary symbols to binary representation.
They work with a family of binary linear error-correcting codes.
Each member of the family is parametrized by an integer $N_\acc$ which represents the number of accepted blocks.
We denote the syndrome function as
${\tt Syn}: \bits^{N_\acc \lceil \log q\rceil }\to \bits^{\qk(N_\acc)}$ 
and syndrome decoding function 
${\tt SynDec}: \bits^{\qk(N_\acc)} \to \bits^{N_\acc \lceil \log q\rceil }$.
For privacy amplification they use a 
collecton of universal hash functions, again parametrized by $N_\acc$, which we write as
{\tt hash}: $\cR\times\bits^{N_\acc \lceil \log q\rceil} \to \bits^{L(N_\acc)}$,
where $\cR$ is the space from which the UHF seed is drawn.

\vskip1mm

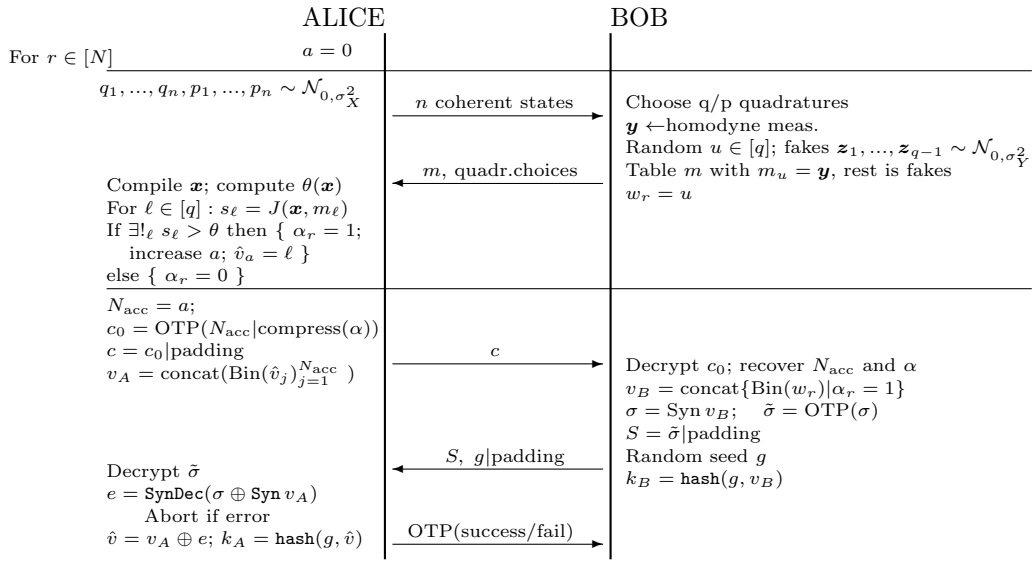
\begin{figure}[h]
%\begin{center}
\setlength{\unitlength}{1mm}
\begin{picture}(150,110)
\put(39,101){ALICE}
\put(80,101){BOB}
\thicklines
\put(50,100){\line(0,-1){70}}
\put(80,100){\line(0,-1){70}} 
\thinlines

\put(39,97){\scriptsize $a=0$}
\put(0,96){\scriptsize For $r\in[N]$}

\put(13,95){\line(1,0){123}}

\put(12,92){\scriptsize $q_1,...,q_n,p_1,...,p_n\sim \cN_{0,\qs_{\tiny X}^2}$}

\put(51,89){\vector(1,0){28}}
\put(54,90){\scriptsize $n$ coherent states}

\put(82,90){\scriptsize Choose q/p quadratures}
\put(82,87){\scriptsize $\vecy\leftarrow$homodyne meas.}
\put(82,84){\scriptsize Random $u\in[q]$; fakes $\vecz_1,...,\vecz_{q-1}\sim\cN_{0,\qs_Y^2}$}
\put(82,81){\scriptsize Table $m$ with $m_u=\vecy$, rest is fakes}
\put(82,78){\scriptsize $w_r=u$}

\put(79,80){\vector(-1,0){28}}
\put(55,81){\scriptsize $m$, quadr.choices}

\put(13,79){\scriptsize Compile $\vecx$; compute $\qy(\vecx)$}
\put(13,76){\scriptsize For $\ell\in[q]: s_\ell=J(\vecx,m_\ell)$}
\put(13,73){\scriptsize If $\exists !_\ell \; s_\ell>\qy$ then \{ $\qa_r=1;$}
\put(16,70){\scriptsize increase $a$; $\hat v_a=\ell$ \} }
\put(13,67){\scriptsize else \{ $\qa_r=0$ \}}

\put(13,66){\line(1,0){123}}

\put(13,63){\scriptsize $N_\acc=a$; }
\put(13,60){\scriptsize $c_0={\rm OTP}(N_\acc|{\rm compress}(\qa))$}
\put(13,57){\scriptsize $c=c_0|{\rm padding}$} 
\put(13,54){\scriptsize $v_A={\rm concat}({\rm Bin}(\hat v_j)_{j=1}^{N_\acc}$ )}

\put(51,56){\vector(1,0){28}}
\put(64,57){\scriptsize $c$}

\put(82,55){\scriptsize Decrypt $c_0$; recover $N_\acc$ and $\qa$}
\put(82,52){\scriptsize $v_B={\rm concat}\{ {\rm Bin}(w_r) | \qa_r=1 \}$}
\put(82,49){\scriptsize $\qs={\rm Syn}\, v_B$; \;\; $\tilde\qs={\rm OTP}(\qs)$}
\put(82,46){\scriptsize $S=\tilde\qs | {\rm padding}$ }
\put(82,43){\scriptsize Random seed $g$}
\put(82,40){\scriptsize $k_B={\tt hash}(g,v_B)$}

\put(79,42){\vector(-1,0){28}}
\put(58,43){\scriptsize $S,\; g|{\rm padding}$}

\put(13,41){\scriptsize Decrypt $\tilde\qs$}
\put(13,38){\scriptsize $e={\tt SynDec}(\qs\oplus {\tt Syn}\, v_A)$}
\put(18,35){\scriptsize Abort if error}
\put(13,32){\scriptsize $\hat v=v_A \oplus e$;  $k_A = {\tt hash}(g,\hat v)$}

\put(51,32){\vector(1,0){28}}
\put(53,33){\scriptsize OTP(success/fail)}

\end{picture}
\vskip-30mm
\caption{
\it 
Protocol steps.
}
\label{fig:steps}
\end{figure}

\underline{Execution}. (See Fig.\,\ref{fig:steps})
\begin{enumerate}[leftmargin=4.5mm,nosep,itemsep=0mm]
\item
Alice initializes a local counter $a=0$.
\item
For $r\in\{1,\ldots,N\}$ Alice and Bob do the following procedure.
\begin{enumerate}[leftmargin=5mm]
\item
{\bf Alice}. Draw $q_1,\ldots,q_n$ and $p_1,\ldots,p_n$ from the distribution $\cN_{0,\qs_X^2}$.
Send $n$ coherent states, where the $i$'th state has displacement $(q_i, p_i)$, 
\item
{\bf Bob}.
For each pulse independently, choose a random quadrature (q or p) to measure.
The measurement outcome is a vector $\vecy\in\RR^n$.
\newline
Draw a random integer $u\in\{1,\ldots,q\}$.
Set $w_r=u$.
Draw random vectors $\vecz_1,\ldots,\vecz_{q-1}$, where $\vecz_j\in\RR^n$ and 
each component of each vector is independently drawn from $\cN_{0,\qs_Y^2}$.
\newline
Construct a matrix $m\in\RR^{q\times n}$ 
of which the rows $\vecm_1,\ldots,\vecm_q$
are the vectors $\vecz_1,\ldots,\vecz_{q-1}$, but
with $\vecy$ inserted in the $u$'th position.\footnote{
I.e. $\vecm_j=\vecz_j$ for $j<u$; $\vecm_u=\vecy$; and $\vecm_j=\vecz_{j-1}$ for $j>u$.
}
Send $m$ and the quadrature choices.
\item
{\bf Alice}.
Collect the correct quadrature component from each pair $(q_i, p_i)$; the result is a vector $\vecx\in\RR^n$.
Compute threshold $\qy(\vecx^2)$.
For each $\ell\in\{1,\ldots,q\}$ compute score $s_\ell = J(\vecx, \vecm_\ell)$, where
\be
    J(\vecx, \vecm) \isdef \frac{ \frac{\vecm}{\qs_Y}\cdot \frac{\vecx}{\qs_X\sqrt n}
    +\frac{1}2\sqrt{\frac{n\qe}{1+\qe}}(1-\frac{\vecm^2}{n\qs_Y^2})  }
    {\sqrt{  \frac{\vecx^2}{n\qs_X^2} +\frac12 \frac{\qe}{1+\qe}  }   }.
\label{defJ}
\ee
\newline
If there exists exactly one index $\ell$ such that $s_\ell>\qy(\vecx^2)$ then \{set $\qa_r=1$; 
increase the counter~$a$;
set $\hat v_a=\ell$
\},
otherwise \{set $\qa_r=0$\}.
\end{enumerate}
\item
{\bf Alice}. Set $\Nacc=a$. 
Create a binary string $v_A$ by concatenation of the strings ${\rm Bin}(\hat v_j)$, with $j$ running from 1 to $\Nacc$.
Prepare a short encoding\footnote{
The size of $\cA$ is $\log{N \choose {\Nacc}}$ bits.
}
$\cA$ of the acccept/reject indicator string $\qa\in\bits^N$;
Compute ciphertext $c_0={\tt Encrypt}(N_\acc| \cA)$, a one-time pad encryption of the concatenation $N_\acc|\cA$.
Set $c=c_0|{\rm padding}$, where the padding is random and serves to create a fixed-length string~$c$.
Send $c$.
\item
{\bf Bob}. 
Recover $N_\acc$ and $\cA$ by decrypting~$c_0$.
Recover $\qa$ from $\cA$.
Create a binary string $v_B$ by concatenating $\{ {\rm Bin}(w_r) | \qa_r=1 \}$.
Compute $\qs={\tt Syn}(v_B)$.
Draw a random seed~$g$.
Compute the key $k_B={\tt hash}(g,v_B)$.
Compute ciphertext $\tilde\qs$, being a one-time pad encryption of~$\qs$.
Set $S=\tilde\qs|{\rm padding}$, where the padding is random and serves to create a fixed-length string.
Send $g$ and~$S$.
Here the seed $g$ is also random-padded to fixed length.
\item
{\bf Alice}. 
Recover $\qs$ by decrypting $\tilde\qs$.
Compute the error pattern $e={\tt SynDec}(\qs\oplus {\tt Syn}\, v_A)$.
If  ${\tt SynDec}$ gives an error, abort.
Otherwise compute $\hat v=v_A \oplus e$
and key $k_A = {\tt hash}(g,\hat v)$.
Send an encrypted success/fail bit to Bob.
\end{enumerate}

Note that the length of $\cA$, $\qs$ and $g$ depends on $N_\acc$. 
The padding in steps 3 and 4 serves to hide~$N_\acc$.
When Bob receives $c$, he recovers $N_\acc$ from the first $\log N$ bits of the ciphertext;
once he knows $N_\acc$ he knows the size of $\cA$ and he ignores the padding.
When Alice receives $S$, she already knows $N_\acc$ and hence the length of $\qs$. 
Similarly, she knowns the length of $g$ beforehand.

%-------------------------------------------------------------------------------
\subsection{Motivation of the score function}
Our choice of score function is motivated by the fact that $J(\vecx,\vecm_\ell)$
 is equivalent to the Neyman-Pearson score for the hypothesis test with null hypothesis
``$u=\ell $'' and competing hypothesis ``$u\neq \ell$''.
This is shown in Appendix~\ref{app:scoreNP}.
For choosing between more than two hypotheses there is no known optimal algorithm.
Therefore we have settled for a possibly sub-optimal solution by splitting up Alice's $q$-fold hypothesis test  
into $q$ binary tests which we know how to handle.

%-------------------------------------------------------------------------------
\subsection{The `decoupling' property}
\label{sec:decoupling}

The classical messages visible to Eve satisfy the `decoupling' property,
i.e.\;all the messages are independent of $u$,
which then allows us to invoke Corollary~\ref{corol:decoupling}.
Eve sees the quadrature choices, all the tables $m$, the ciphertext $c$, the ciphertext $S$, the seed $g$,
and the encrypted success/fail bit.
The index $u$ is obviously independent of the quadrature choices. 
Furthermore, all the ciphertexts are made by OTP-ing, which makes them into independent variables.
The seed $g$ is chosen independently. 
Finally, for the tables it holds that $I(M;U)=H(M)-H(M|U)$, and $H(M|U)=H(M)$.
The latter holds because the fake rows follow the same statistics as the row that contains~$\vecY$.

%----------------------------------------------------------------------------

%===========================================
\section{Statistical properties of the scores; error probabilities}
\label{sec:stat}

We briefly analyse the statistics of the scores~$s_\ell$ (\ref{defJ}).
We distinguish between the case where $\vecm_\ell$ is a random vector independent of $\vecx$
and the case where $\vecm_\ell = \vecy$.
In the former case we write $\vecm_\ell=\vecZ$.
In the latter case we write $\vecy=\vecx\sqrt T+\vecN$, where the noise components $N_i$ are gaussian
with zero mean and variance $\qs_{Y|X}^2$.
In terms of the $\qe$ parameter this can be written as 
\be
    \frac{\vecy}{\qs_Y} = \frac{\frac\vecx{\qs_X}\sqrt\qe +\vecN/\qs_{Y|X}}{\sqrt{1+\qe}}.
\ee

The score function $J$ has been engineered such that, at given $\vecx$, the scores of the fake table entries have zero average and unit variance,
\be
    \EE_Z J(\vecx,\vecZ) = 0  \quad\quad
    {\rm Var}_Z  J(\vecx,\vecZ) = 1.
\ee
At given $\vecx$, the mean and variance of the correct-symbol score $J(\vecx,\vecY)$ are given by
\be
    \EE_{Y|x} J(\vecx,\vecY) = 
    \frac{\sqrt{n\qe}}{(1+\qe)^{3/2}}
    \frac{   \frac{\vecx^2}{n\qs_X^2 }(1+\frac\qe2) +\frac\qe2 }
    { \sqrt{  \frac{\vecx^2}{n\qs_X^2} +\frac12 \frac{\qe}{1+\qe}  }   }
    = \sqrt{n\qe}[\sqrt{\frac{\vecx^2}{n\qs_X^2}} + \cO(\qe)]
\label{meanJu}
\ee
\be
    {\rm Var}_{Y|x}J(\vecx,\vecY) = \frac1{(1+\qe)^3} \frac{ \frac{\vecx^2}{n\qs_X^2} +\frac\qe2}
    { \frac{\vecx^2}{n\qs_X^2} +\frac12 \frac{\qe}{1+\qe}}
    =1+\cO(\qe).
\ee
We further examine the distribution of the fake-entry score $J(\vecx,\vecZ)$.
We introduce the notation
$\frac{\vecZ}{\qs_Y}=\qz_\parallel \frac{\vecx}{|\vecx|}+\vecqz_\perp$,
where $\qz_\parallel$ represents the part of $\frac\vecZ{\qs_Y}$ parallel to $\vecx$
and $\vecqz_\perp$ the part perpendicular to $\vecx$.
This gives 
$\frac{\vecZ}{\qs_Y}\cdot \frac{\vecx}{\qs_X\sqrt n} = \qz_\parallel \frac{|\vecx|}{\qs_X\sqrt n}$.
(One can rotate the basis of the $n$-dimensional vector space and let the first basis vector point in the
direction of $\vecx$. Then $\qz_\parallel$ is the component in the first basis direction.)
Note that $\qz_\parallel$ is normal-distributed.

Note that $\vecqz_\perp^2$ is a scalar variable with a $\chi^2_{n-1}$ distribution.
With this notation, the fake-entry score is expressed as
\be
    J(\vecx, \vecZ) = 
    \frac{ \qz_\parallel \frac{|\vecx|}{\qs_X\sqrt n}
    +\frac{1}2\sqrt{\frac{n\qe}{1+\qe}}(1-\frac{\qz_\parallel^2 + \vecqz_\perp^2}n)  }
    {\sqrt{  \frac{\vecx^2}{n\qs_X^2} +\frac12 \frac{\qe}{1+\qe}  }   }
    = \qz_\parallel + \cO(\sqrt\qe)+\cO(\frac1{\sqrt n}).
\label{Jzetapar}
\ee
Here we have used that 
$\frac{\vecx^2}{n\qs_X^2}=1+\cO(1/\sqrt n)$, 
$\qz_\parallel^2=1+\cO(1)$ and
$\vecqz_\perp^2 = n-1+\cO(\sqrt n)$ with overwhelming probability.
From (\ref{Jzetapar}) we observe that the fake-entry score $J(\vecx,\vecZ)$ consists almost entirely
of the normal-distributed variable $\qz_\parallel$.

\begin{figure}[h!]
\begin{center}
\begin{picture}(140,85)
\setlength{\unitlength}{1mm}
\put(0,-6){\includegraphics[width=.35\textwidth]{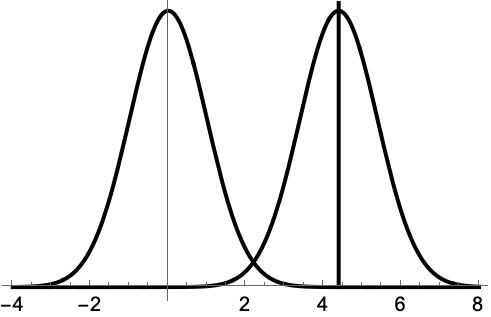}}
\put(17,-6){$0$}
\put(31,30){$\frac{\sqrt{2\ln q}}{\sqrt{1+\qg}}$}
\put(52,-2){score}
\put(42,18){\scriptsize pdf of the}
\put(42,15){\scriptsize correct-entry}
\put(42,12){\scriptsize score}
\put(0,18){\scriptsize pdf of}
\put(0,15){\scriptsize fake-entry}
\put(0,12){\scriptsize score}
\end{picture}
\end{center}
%\vskip4mm
\caption{\it
Probability density function of the score for a fake table entry and for the correct $(u$'th) table entry.
Contributions of order $\cO(\sqrt\qe)$ are neglected.
The pdfs are almost Gaussian and have variance~1. 
The plot is for $q=2^{10}$, $\qg=-0.28$
}
\label{fig:bulten}
\end{figure}
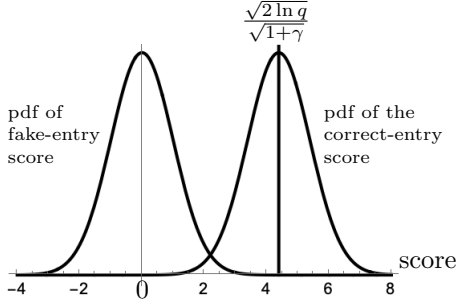

The amount of mutual information between Alice and Bob contained in $n$ pulses is given by
$I(\vecX;\vecY)=nI(X_i;Y_i)=n\cdot\frac12\log(1+\qe)$.
Ideally, $\log q$ is set equal to $I(\vecX;\vecY)$, so that capacity is achieved.
We will introduce a parameter $\qg$ that encodes how far away from capacity we operate,
\be
    \log q = (1+\qg) n \cdot \frac12\log(1+\qe).
\label{introgamma}
\ee
$\qg<0$ means operating below capacity.
We will write $n$ expressed in $\qe,q,\qg$, i.e.
$n=\frac{2\ln q}{(1+\qg)\ln(1+\qe)}$.
With this notation, the mean (\ref{meanJu}) is expressed as
\be
    \EE_{Y|x} J(\vecx,\vecY) =  \frac{\sqrt{2\ln q}} {\sqrt{1+\qg}}+\cO(\sqrt\qe).
\ee
We also introduce notation $\frac\vecN{\qs_{Y|X}}=\nu_\parallel \frac{\vecx}{|\vecx|}+\vecnu_\perp$,
where $\nu_\parallel$ is normal-distributed.
\bea
	J(\vecx,\vecY) &=& 
	\frac1{   \sqrt{\frac{\vecx^2}{n\qs_X^2}(1+\qe)+\qe/2}  }\Big[
	\frac{\vecx^2}{n\qs_X^2}\sqrt{n\qe}\frac{1+\qe/2}{1+\qe}
	+\nu_\parallel \sqrt{\frac{\vecx^2}{n\qs_X^2}}\frac{1+\qe/2}{1+\qe}
	+\frac{\sqrt{n\qe}}{2}(1-\frac{\nu_\parallel^2+\vecnu_\perp^2}{n(1+\qe)})
	\Big]
	\nn\\ &=&
	\frac{\sqrt{2\ln q}} {\sqrt{1+\qg}} + \nu_\parallel +\cO(\sqrt\qe).
\label{JxYstochast}
\eea
Motivated by (\ref{Jzetapar}) and (\ref{JxYstochast}) we model all the scores as
gaussian random variables, neglecting terms of order $\sqrt\qe$.
(See Fig.\,\ref{fig:bulten} for a visualisation.)

\vskip3mm

\underline{Error probabilities}.\\
We place the threshold $\qy$ near the average true-entry score, $\EE_{Y|x} J(\vecx,\vecY)$.
\be
	\qy = \frac{\sqrt{2\ln q}} {\sqrt{1+\qg}} + \qd.
\label{introdelta}
\ee
With the gaussian approximation we compute the True Accept (TA) and False Accept (FA) probabilities as follows,
\bea
	P_{\rm TA}= [\qF(\qy)]^{q-1}[1-\qF(\qd)]
	\quad ; \quad
	P_{\rm FA} = (q-1)[\qF(\qy)]^{q-2} \,[1-\qF(\qy)]\, \qF(\qd),
\eea
where $\qF$ is the cumulative distribution functon of the normal distribution.
A True Accept occurs when all $q-1$ fake-entry scores lie below $\qy$ and the true-entry score lies above~$\qy$.
A False Accept occurs when $q-2$ fake-entry scores lie below $\qy$, one fake-entry score above $\qy$,
and the true-entry score below~$\qy$.
The factor $q-1$ occurs because there are $q-1$ choices which of the fakes has a high score.

For the total accept probability we write $P_\acc=P_{\rm TA}+P_{\rm FA}$.
Each accept gives rise to a $q$-ary symbol being kept as part of the raw key.
If Bob's reconstructed entry does not equal Alice's $u$, we say that a {\em symbol error} has occurred.
The symbol error probability $\qs$ is given by
\be
	\qs= \frac{P_{\rm FA}}{P_{\rm FA}+P_{\rm TA}} 
	= \Big[1+ \frac1{q-1}\frac{\qF(\qy)}{1-\qF(\qy)} \frac{1-\qF(\qd)}{\qF(\qd)}  \Big]^{-1}.
\label{SER}
\ee  
A symbol error leads to bit errors in approximately 50\% of the bits in the binarisation of the symbol.

\begin{lemma}
Let $q$ be a power of~$2$.
The bit error rate (BER) follows from the symbol error rate as
\be
    {\rm BER} = \frac{\qs}{2} \cdot\frac{q}{q-1}.
\label{BERsigma}
\ee
\end{lemma}
\underline{\it Proof:}
Let $u$ be Alice's $q$-ary symbol, and let $\hat u$ be Bob's reconstruction of~$u$.
The binarisation is ${\rm Bin}(u)\in\bits^{\log q}$.
For any bit position index $j$ it holds that
$\pr[{\rm Bin}(\hat u)_j \neq {\rm Bin}(u)_j ] $
$= \pr[\hat u\neq u] \frac{ |\{\hat u:\; {\rm Bin}(\hat u)_j\neq  {\rm Bin}(u)_j\}| }{|\{ \hat u : \; \hat u\neq u \}|}$
$=
    \qs \frac{q/2}{q-1}
$.
\hfill $\square$

%============================================================================
\section{Asymptotic Secret Key Ratio}
\label{sec:SKR}

As our protocol (and the subsequent key combination procedure) hides $N_\acc$
and the individual accept/reject decisions from the adversary,
we can use Eve's {\em gaussian} states $\qr^E$ and $\qr^E_y$ without further conditioning.
The classical messages accessible to Eve satisfy the `decoupling' property (Section~\ref{sec:decoupling}),
which allows us to invoke Corollary~\ref{corol:decoupling}.
From Corollary~\ref{corol:decoupling} together with being allowed to use the standard formula for the leakage we get
the following expression,
\be
	I(W_1\ldots W_{N_\acc}; E, {\rm transcript} ) \leq N_\acc n I(E_i;Y_i).
\ee
The size of the raw key (before privacy amplification) shared between Alice and Bob is 
$N_\acc \log q$.
However, because of the nonzero symbol error rate a second layer of error correction is needed,
e.g. in the form of a binary linear code or a nonbinary burst correction code.
Since we are studying the asymptotics, we will assume that this code is operating close to the
capacity $1-h({\rm BER})$.
Taking this into account, the size of the syndrome $\qs$ is 
$h({\rm BER}) N_\acc\log q $.

In the calculation of the key ratio we have to subtract the amount of key material that gets
consumed in the encryption of the integer $N_\acc$, the compressed acceptance vector $\qa$, the syndrome~$\qs$,
and the final success/fail bit.
The $N_\acc$ is expressed in at most $\log N$ bits.
The size of the compressed $\qa$ is $\log{N\choose N_\acc}\leq Nh(P_\acc)$ bits;
the inequality is tight.
Combining these ingredients, we get the following provably safe asymptotic key ratio (${\rm SKR}_\infty$) for our scheme,
\bea
	{\rm SKR}_\infty &  \!\!\! = \!\!\! & 
	\frac1{Nn}\Big[ N_\acc\log q - N_\acc n I(E_i;Y_i) - \log N - Nh(P_\acc) - h({\rm BER}) N_\acc\log q -1\Big]
	\nn\\ & \!\!\! = \!\!\! &
	P_\acc\Big[ (1+\qg)\frac12\log(1+\qe)\big\{1-h(\frac\qs2\cdot \frac{q}{q-1}) -\frac{h(P_\acc)/P_\acc}{\log q} \big\}
    -I(E_i; Y_i)\Big] - \frac{\log 2N}{Nn}.
    \quad\quad
\label{schemeSKR}
\eea
Here we used (\ref{introgamma}) to write $n$ in terms of $q,\qe,\qg$,
and we substituted (\ref{BERsigma}).
Asymptotically the term $\frac{\log 2N}{Nn}$ is negligible.

In Table~\ref{table:SKRDW} we show optimized key ratios as a function of $q$, at $T=10^{-6}$, $\xi=10^{-5}$.
When $T,\xi,q$ are fixed, optimization consists of finding $\qs_X^2,\qg,\qd$ such that the SKR is maximized.
The table lists the SKR normalized with respect to the Devetak-Winter value; 
this has the advantage that we do not have to present a table for different values of~$T$.
The results at $T=10^{-6},\xi=10^{-5}$ are representative for $T\ll 1,\xi\ll 1$.
Without showing details we mention that the numbers in Table~\ref{table:SKRDW} hardly change when $T$ is varied.

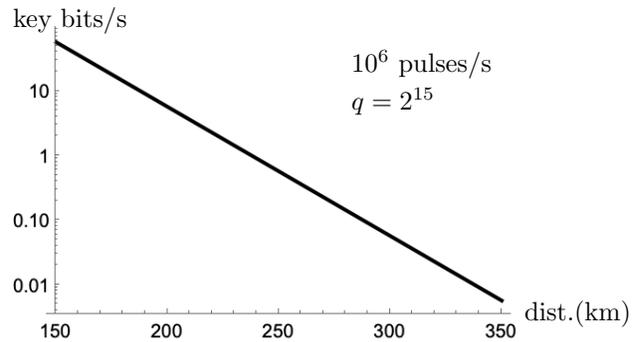
\begin{figure}[ht]
\begin{center}
\begin{picture}(140,105)
\setlength{\unitlength}{1mm}
\put(0,-6){\includegraphics[width=.45\textwidth]{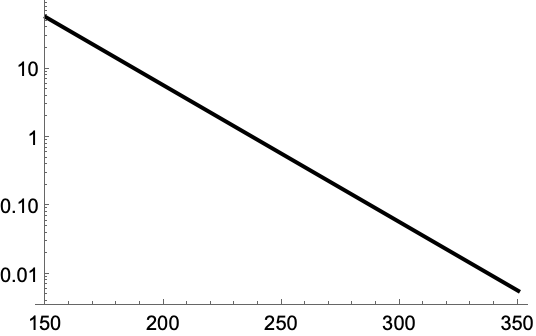}}
\put(68,-3){dist.(km)}
\put(0,36){key bits/s}
\put(45,30){$10^6$ pulses/s}
\put(45,25){$q=2^{15}$}
\end{picture}
\end{center}
%\vskip4mm
\caption{\it
Secret key rate in bits per second, according to Table~\ref{table:SKRDW},
at $q=2^{15}$, given a pulse rate of $10^6$ pulses per second.
The loss is taken to be 0.2 dB/km. Excess noise is neglected.
}
\label{fig:dist}
\end{figure}

Fig.\;\ref{fig:dist} shows an example of the key rate in bits per second, at $q=2^{15}$, to illustrate the orders of magnitude
at a realtime implementable value of~$q$.

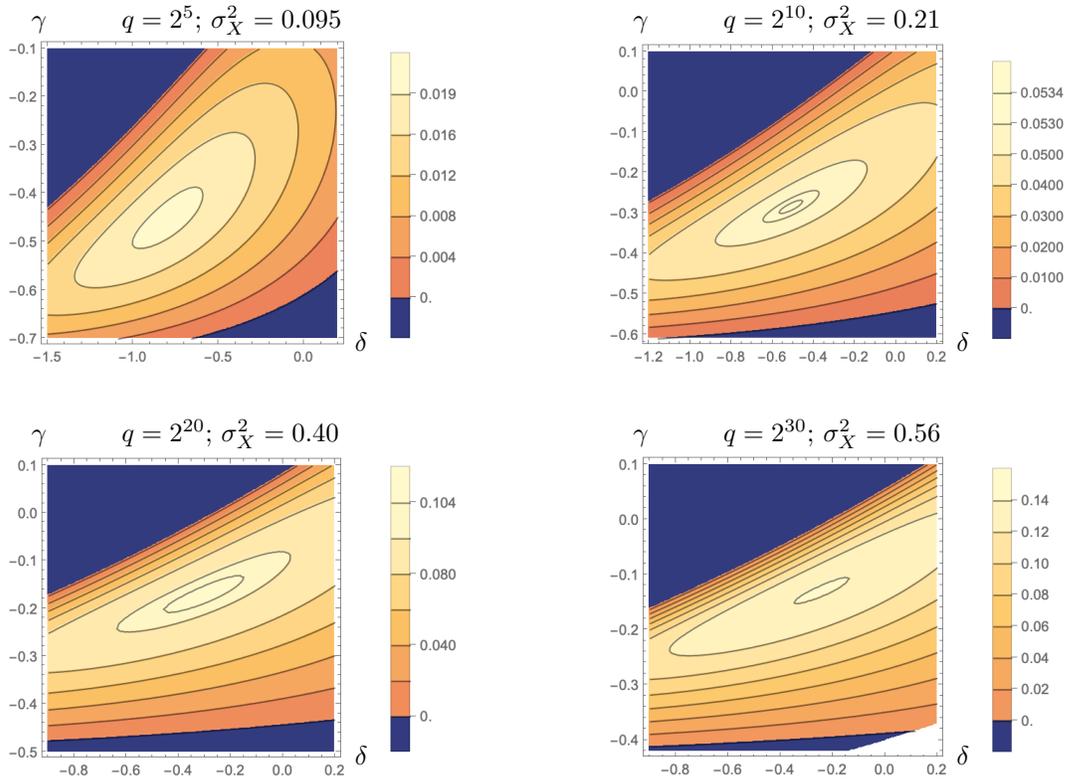
\begin{figure}[ht]
\setlength{\unitlength}{1mm}
\begin{picture}(150,110)

\put(0,55){\includegraphics[width=.3\textwidth]{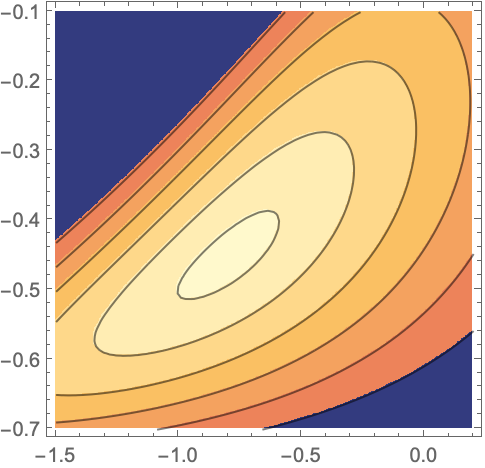}}
\put(46,57){$\qd$}
\put(3,100){$\qg$}
\put(15,100){$q=2^5$; $\qs_X^2=0.095$}
\put(50,55){\includegraphics[width=.065\textwidth]{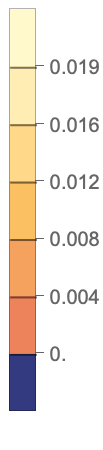}}

\put(80,55){\includegraphics[width=.3\textwidth]{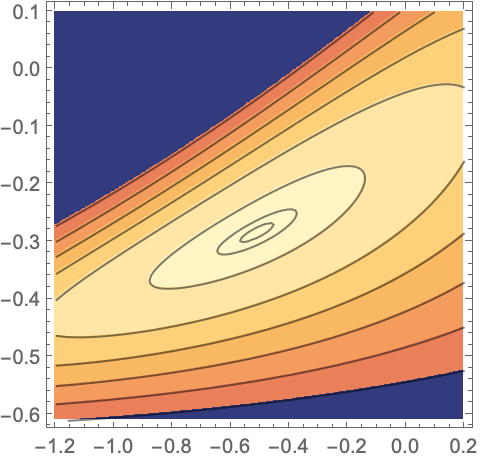}}
\put(126,57){$\qd$}
\put(83,100){$\qg$}
\put(95,100){$q=2^{10}$; $\qs_X^2=0.21$}
\put(130,55){\includegraphics[width=.07\textwidth]{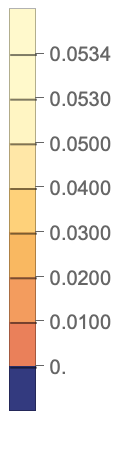}}

\put(0,0){\includegraphics[width=.3\textwidth]{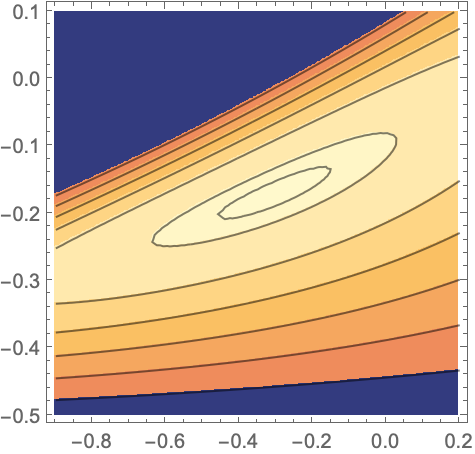}}
\put(46,2){$\qd$}
\put(3,45){$\qg$}
\put(15,45){$q=2^{20}$; $\qs_X^2=0.40$}
\put(50,0){\includegraphics[width=.065\textwidth]{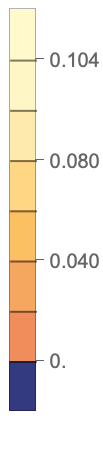}}

\put(80,0){\includegraphics[width=.3\textwidth]{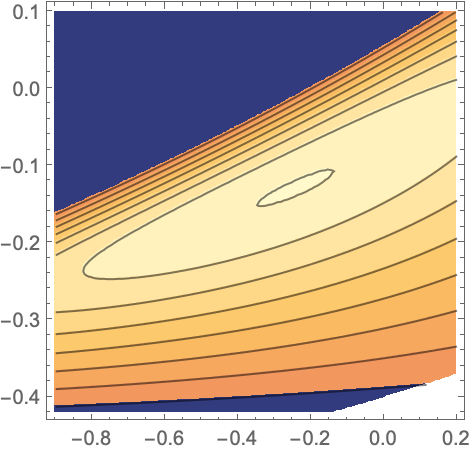}}
\put(126,2){$\qd$}
\put(83,45){$\qg$}
\put(95,45){$q=2^{30}$; $\qs_X^2=0.56$}
\put(130,0){\includegraphics[width=.057\textwidth]{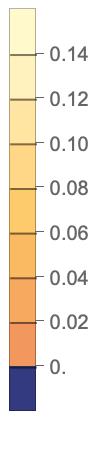}}

\end{picture}
%\vskip-30mm
\caption{
\it 
SKR$_\infty$/DW as a function of $\qg$ and $\qd$, for various combinations of $q$ and $\qs_X^2$ (see Table~\ref{table:SKRDW}).
$T=10^{-6}$, $\xi=10^{-5}$.
}
\label{fig:SKR}
\end{figure}

Fig.\;\ref{fig:SKR} zooms in on the SKR `landscape' as a function of $\qg$ and $\qd$ at optimal $\qs_X^2$. 
We observe the following trends.
(i) 
The SKR is several percent of the DW value, which is a good result given the low~$T$. 
The SKR slowly increases with growing~$q$.
From the listed SKR/$\triangle I$ values, we see that this is partly due to the low value of $\qs_X^2$, which lowers the achievable rate,
and partly due to the fact that the scheme performs at 13\% to 29\% of $\triangle I$.
(ii)
The average photon number is below~1, even at very large $q$.
(iii)
The parameters $\qg$ and $\delta$ are negative, getting closer to zero as $q$ increases.
Negative $\qg$ moves the threshold $\qy$ to the right, but negative $\qd$ moves it to the left.
The net effect is that the symbol error rate is kept low, around 1\% or 2\%, while 
$P_\acc$ ranges from 59\% to 74\%.
We note that our assumption about the asymptotic size of the syndrome is well justified at low~BER.

We briefly comment on a scheme variant without thresholding.
That variant always accepts, and is hence very simple to analyze.
It has symbol error rate $\qo=1-\EE_{\ql\sim\cN_{01}} [\qF(\frac{\sqrt{2\ln q}}{\sqrt{1+\qg}}+\ql)]^{q-1}$ and
asymptotic key ratio $(1+\qg)\frac12\log(1+\qe)[1-h(\frac\qo2\frac{q}{q-1})] - I(E_i;Y_i)$.
At large $q$ the $\qo$ is small enough to justify the asymptotic expression $1-h(..)$, but the key ratio is worse than
for the thresholded scheme.
At small $q$ the asymptotic key ratio is slightly better than for the thresholded scheme, but the
$\qo$ grows, making it difficult to justify the $1-h(..)$ expression.

\begin{table}[h!]
\caption{\it
Values of $\qs_X^2$, $\qg$, $\qd$ that optimize SKR$_\infty$, listed for several values of $q$.
The channel parameters are $T=10^{-6}$ and $\xi=10^{-5}$.
The fraction SKR$_\infty$/DW is the secret key ratio (\ref{schemeSKR}) divided by the Devetak-Winter value $T/(2\ln 2)$.
In (\ref{schemeSKR}) we use (\ref{introdelta}), (\ref{SER}), (\ref{leakFull}).
The accept probability and symbol error rate $\qs$ at the optimum are also listed.
``$\triangle I$'' stands for $I(X_i;Y_i)-I(E_i;Y_i)$ at the listed $\qs_X^2$.
}
\label{table:SKRDW}
\begin{center}
\begin{tabular}{|c||c|c|c||c|c|c|c|}
\hline
$q$ & $\qs_X^2$ \vphantom{$M^{M^M}$} & $\qg$ & $\qd$ & SKR$_\infty$/DW & $P_\acc$ & $\qs$ & SKR$_\infty/\triangle I$
\\ \hline
$2^{5}$ \vphantom{$M^{M^M}$}& $0.095$ & $-0.45$ & $-0.78$ & $0.019$ & $0.74$ & $0.024$ & $0.13$ 
\\ \hline
$2^{10}$ \vphantom{$M^{M^M}$}& $0.21$ & $-0.28$ & $-0.50$ & $0.053$ & $0.67$ & $0.022$ & $0.20$
\\ \hline
$2^{15}$ \vphantom{$M^{M^M}$}& $0.31$ & $-0.21$ & $-0.40$ & $0.082$ & $0.64$ & $0.020$  & $0.24$
\\ \hline
$2^{20}$ \vphantom{$M^{M^M}$}& $0.40$ & $-0.17$ & $-0.30$ & $0.11$ & $0.61$ & $0.014$ & $0.26$
\\ \hline
$2^{30}$ \vphantom{$M^{M^M}$}& $0.56$ & $-0.13$ & $-0.24$ & $0.14$ & $0.59$ & $0.0097$ & $0.29$
\\ \hline
\end{tabular}
\end{center}
\end{table}

%\clearpage

%\clearpage

\clearpage

%===========================================
\section{Implementation aspects}
\label{sec:imp}

%---------------------------------------------------------------------------------
\subsection{Decoding}
\label{sec:decoding}

In CVQKD the main computational bottleneck is the error-correction decoding on Alice's side.
Our scheme does not alter that fact, but it allows for massive parallelization: all the scores $(s_\ell)_{\ell=1}^q$
can be computed independently, and even within $s_\ell=J(\vecx,\vecm_\ell)$ all the terms within the inner products
$\vecx\cdot\vecm_\ell=\sum_{i=1}^n x_i m_{\ell i}$ and $\vecm_\ell^2=\sum_{i=1}^n (m_{\ell i})^2$
can be computed in parallel.
(Note that $\vecx^2$ can be precomputed.)

In a real-world CVQKD system, Alice has to do the decoding realtime.
This means that the time spent on computations pertaining to $Nn$ pulses
should not take longer than the time that it takes to send those $Nn$ pulses.
(Latency is not important. A time lag can be tolerated.) 
%Let us write $J(\vecx,\vecm)=\sum_{i=1}^n $
Hence, in the time interval between two pulses, $P_\acc q$ basic multiplications $x_i m_{\ell i}$ and $(m_{\ell i})^2$ must be performed,
plus additions.\footnote{
This is not counting the outer error correction, which we consider to be `easy' given the low BER.
}
For example, if Alice fires  $10^6$ pulses per second, and she tries to implement our scheme with $q=2^{15}$,
she will need to do $P_\acc\cdot 2^{15} \approx 2.1\cdot 10^4$ multiplications in 1 microsecond.
On a 5~GHz processor that means 5000 clock cycles, which implies around 4 multiplications per clock cycle.
With a handful of of cores that task should be achievable.

We mention a number of options that can make the implementation faster.
\begin{itemize}[leftmargin=4.5mm,nosep,itemsep=0mm]
\item
In the computation of $J(\vecx,\vecm)$, initially ignore the contribution from $1-\frac{\vecm^2}{n\qs_Y^2}$, since
it is often negligible.
Compute this contribution only if the $\vecx\cdot \vecm$ part of the score is close to the threshold~$\qy$.
\item
If two scores have already been found that exceed the threshold, there is no need to compute the leftover scores. 
\item
As the index $i$ evolves from $1$ to $n$, the scores `grow' over time.
It can be advantageous to postpone the full computation of scores that at an early stage (e.g. $i=n/4$ or $n/2$) 
start out negative, since they will at the end exceed $\qy$ only with low probability.
It may even be possible to discard them.
\item
The scheme as described in Section~\ref{sec:steps}
works with reals, but of course we have to discretize.
A good way to discretize continuous variables while retaining a lot of information
is to use quantiles, in such a way that
the binary representation has a uniform distribution.
For instance, the component $z_i/\qs_Y$ of a fake table entry is normal-distributed,
and its discretization to a $b$-bit binary string is given by an integer in $\{0,\ldots,2^b-1\}$
determined as $\lfloor 2^b \qF(z_i/\qs_Y) \rfloor$.
The same is done with $x_i/\qs_X$ and $y_i/\qs_Y$.
Note that the discretization does not impact the security analysis.
Eq.(\ref{schemeSKR}) remains valid, but the symbol error rate $\qs$ may become worse due to the loss 
of information.

After the discretization, it will not be necessary to actually {\em compute} multiplications
$\frac{x_i}{\qs_X}\frac{m_{\ell i}}{\qs_Y}$.
Instead, we can precompute a $2^b \times 2^b$ LookUp Table (LUT) for all combinations of the discretized inputs
$\frac{x_i}{\qs_X}$ and $\frac{m_i}{\qs_Y}$,
and look up values from the LUT.
Optionally, for small $b$, computing a score $s_\ell$ could then be implemented as
(i)
aggregating tallies $(t_{jk})_{j,k\in\{0,\ldots,2^b-1\}}$ that count how often input combination $(j,k)$ occurs;
(ii)
then compute $\sum_{jk}t_{jk}{\rm LUT}_{jk}$.
\end{itemize}
Ideally, one would have a large number of mini-cores, each of which 
receives inputs in a pipelined way and produces either LUT lookups or tallies as defined above.
At a downstream layer the results are aggregated and decisions are made which table entries to discontinue.

%----------------------------------------------------------------------------------
\subsection{Communication}
\label{sec:PRNG}

Our scheme requires much more communication than other reconciliation methods.
For each pulse that contributes to the final key,
Bob needs to send $q$ numbers to Alice. 
At large $q$ this becomes expensive or even infeasible.
We propose a protocol modification that replaces the truly random fakes by
pseudorandom sequences.

Let the `tilde' notation denote $b$-bit discretization as above.
%$\tilde y_i$ be the discretisation of $y_i$ as described above.
Let $\qj(\cdot,\cdot,\cdot)$ be a function that takes as input a seed and two integers,
and outputs a pseudorandom sequence.
Steps 2b,2c of the protocol are modified as follows.
\begin{itemize}[leftmargin=4.5mm,itemsep=0mm]
\item
Bob draws a random seed~$\qt$.
For $j\in\{1,\ldots,k\}$ he computes $a_j=\qj(\qt,u,j)$, and $a=a_1|\cdots|a_k$.
Here $k$ is an integer that is tuned such that $a$ has the same length as $\tilde y$.
Bob computes $\mu=\tilde y \oplus a$, where $\oplus$ stands for bitwise XOR.
He sends $\qt,\mu$.
\item
For $\ell\in[q]$, Alice computes scores $J(\vecx,\vecm_\ell)$, where the vector $\vecm_\ell$
is obtained as
$\tilde \vecm_\ell = \mu\oplus(g_{\ell 1}| \cdots|g_{\ell k})$, with $g_{\ell j}=\qj(\qt,\ell,j)$.
\end{itemize}
At $\ell=u$ Alice retrieves $\tilde\vecm_u = \tilde y$.
At $\ell\neq u$ she gets
$\tilde y$ xor-ed with two combined pseudorandom sequences.

This construction resembles schemes that XOR the $\tilde y$, or some other representation of $y$,
with a randomly chosen codeword from a linear error correction code.
In our case, however, the code is entirely unstructured and allows for
parallel decoding. 
In particular, the pieces $\qj(\qt,\ell,j)$ can be computed in parallel for all $j$, which implies that 
the corresponding contributions to the score $s_\ell$
can be done in parallel.
Note that spinal codes
\cite{PBS2011,BIPS2012,WLMLYH2020}
(another type of pseudorandom code)
do not have this property; they need to generate a codeword sequentially.

Note that $I(U; \qt\mu)=0$,\footnote{
$t$ is independent of $u$. Furthermore, the $\tilde y$ is uniform and 
serves as a one-time pad, completely hiding the $\qj(\qt,u,j)$.
} which implies that we are still allowed to use Corollary~\ref{corol:decoupling},
and the bound on the leakage, $I(U;\qt\mu E)\leq I(E;Y)$, remains valid.
One must still be careful with the choice of pseudorandom function $\qj$,
since 
bad pseudorandomness introduces patterns into the sequences $\qj(\qt,\ell,j)$,
which means that the vectors $\vecm_l\in \RR^n$ are not evenly spread out, leading to inefficiencies
in the decoding.
A good choice might be to use unbiased sets \cite{ABNNR92,AGHP92,NaorNaor93,AmbainisSmith2004},
pairwise independent hash functions, or
or a fast stream cipher, e.g. ChaCha20 \cite{ChaCha20}.
However, for coding efficiency it is probably not required to have 
cryptography-strength pseudorandomness.
We mention that PRNG algorithms exist which produce multiple bytes 
of pseudorandom output per clock cycle
\cite{gjrand,digicortex,xoshiro,shishua} on a CPU.

%===========================================
\section{Discussion}
\label{sec:discuss}

The main result of this paper
is the key ratio (\ref{schemeSKR}) obtained from the random codebook with MAP score thresholding 
and encryption of accept/reject decisions.
We mention again that many works on reconciliation in CVQKD do not
fully take into account that at nonzero block rejection rate 
%(often called Frame Error Rate)
the relevant part of Eve's state gets conditioned on the `accept' outcome,
which negatively affects the key ratio analysis.
Hence it is difficult to gauge the accuracy of
the claimed key rates and \%DW values listed in Table~\ref{tab:cvqkd_ec_long_distance},
except when the frame rejection rate is close to zero, as in \cite{hajomer100km}. 

The pseudorandom scheme variant presented in Section~\ref{sec:PRNG} is the one that is
most easy to implement
and should be seen as the `default' one that we propose.
It strongly resembles existing schemes which xor a randomly chosen codeword from an error-correcting code into
Bob's measurement~$Y$.
The main operational difference is in the choice of code: in our case the decoding is massively parallelizable and works even at very low
SNR.
The complexity of the decoding procedure increases with worsening SNR in a very friendly way, in the sense that
the sequence length $n$ is proportional to $1/\qe$, and the decoding workload (at fixed $q$) is linear in~$n$. 
As the example in Section~\ref{sec:decoding} shows, 
we expect the $q=2^{15}$ scheme to be implementable by sheer `brute force', on a small number of multicore CPUs,
without any optimizations such as pruning.
It is left for future work to determine how much can be gained from aggressive pruning
and optimized parallel implementation on FPGA.

The obtained SKR values are competitive
(see Table~\ref{tab:cvqkd_ec_long_distance})
given that we operate at extremely low~$T$.
At $q=2^{15}$ our scheme attains 8\% of the Devetak-Winter value;
this slowly increases with larger~$q$.
It is difficult to get closer to the DW value for the following reason.
The distance between the two Gaussian curves is
$\sqrt{2\ln q}$ is we set $\qg=0$, i.e.\;if we try to operate at Shannon capacity $I(X_i;Y_i)$.
At `reasonable' values of~$q$, this distance is not sufficient to get a good separation between the $q-1$ fake scores
and the true-entry score.
Therefore we take negative~$\qg$; 
that lowers the efficiency of the code; 
the low efficiency forces us to work with small modulation variance $\qs_X^2$, which
further lowers the SKR.
The only way out of this spiral is to increase $q$, but that has practical limits
because of the realtime constraint.
We emphasize that entirely encrypting all the accept/reject decisions is a drastic measure
that consumes a lot of key material, which hurts the SKR.
We expect that better approaches can be developed.

The analysis of score statistics and secret key rate in Sections \ref{sec:stat} and~\ref{sec:SKR}
is valid, in principle, only for perfectly random fake entries.
However, making the fakes {\em pseudo}random should not strongly affect 
the accept probability $P_\acc$ and symbol error rate $\qs$, which are the two 
statistical quantities that enter into the SKR computation.

In summary, we have shown that 
(1) pseudorandom-codebook error correction
can be a practical option for realtime information reconciliation in long-distance CVQKD,
(2)
even when the security analysis is highly conservative, i.e.\;based on subtracting the length of the OTP that
encrypts the accept/reject decisions.
Here it must be noted that fully hiding the length of the QKD keys is nontrivial, in view of side channels,
and can be realized e.g.\;by always combining the QKD key with a PQC key regardless of the length.

As future work we mention the following topics:
(i)
Generalize to discrete modulation.
(ii)
Study the resolution of the quadrature discretization, in particular determine 
to how many bits $y$ and $z$ should be discretized. 
(iii)
Non-asymptotic (finite~$N$) analysis of the SKR.
(iv)
Study the effect of non-negligible excess noise in the channel.\footnote{
Without showing details, we mention that setting $\xi=10^{-4}$ hardly changes our results.
At $\xi=10^{-3}$ the SKR is noticeably starting to degrade a bit.
At $\xi=10^{-2}$ the SKR at $q=2^{15}$ drops to $0.6$\% of DW; the effect is less dramatic at larger~$q$.
}
(v)
Optimized parallel implementation of the pseudorandom scheme.

\vskip4mm

{\large\bf Acknowledgements}\\
We thank Kadir G\"{u}m\"{u}\c{s}, Jo\~{a}o dos Reis Fraz\~{a}o, and Chigo Okonkwo for useful discussions.
Part of this work was supported by the Dutch NGF Quantum Delta NL KAT-2.

%===========================================
\appendix
%--------------------------------------------------------------------
\section{The leakage (homodyne detection)}
\label{app:leak}

The mutual information $I(E_i;Y_i)$
follows from the symplectic eigenvalues $\nu_1,\nu_2$ of 
Eve's pre-measurement state $\qr^{E_i}$ and  the symplectic eigenvalue $\nu_3$ of the (homodyne)
post-measurement state~$\qr^{A_i}_{y_i}$
\cite{leverrier:tel-00451021}.
\be
	I(E_i;Y_i) = g(\frac{\nu_1-1}2) + g(\frac{\nu_2-1}2) - g(\frac{\nu_3-1}2),
\label{leakFull}
\ee
with $g$ the thermal entropy function  
$g(x)\isdef (x+1)\log_2(x+1)-x\log_2 x$,
and
\be
	\nu^2_{1,2} = \frac{\qD\pm\sqrt{\qD^2-4D}}2
    \quad\quad\quad\quad
    \nu_3^2  = V\Big[ V-\frac{T(V^2-1)}{2\qs_Y^2}    \Big]
\ee
\be
	\qD \isdef  V^2 + [2\qs_Y^2]^2 -2T(V^2-1)
	\quad\quad\quad\quad
	D \isdef \Big[ V\cdot2\qs_Y^2 -T(V^2-1) \Big]^2
\ee
where $V=1+2\qs_X^2$.

%--------------------------------------------------------------------
\section{The score function is equivalent to the likelihood ratio}
\label{app:scoreNP}

For hypothesis tests involving two competing hypotheses,
the {\em likelihood ratio} is known to be the uniformly most powerful test.
Alice knows $\vecx$ and the matrix $m$.
Based on this information, she wants to test the null hypothesis $H_0$: $u=\ell$ versus
competing hypothesis $H_1$: $u\neq \ell$, for some fixed $\ell\in[q]$.

The likelihood ratio is
${\rm LR}=\frac{\pr[H_0 | \vecx,m ]}{\pr[H_1 | \vecx,m ]}$
$=\frac{\pr[H_0 | \vecx,m ]}{ 1-\pr[H_0 | \vecx,m ] }$.
Since $\frac x{1-x}$ is an increasing function, thresholding 
$\pr[H_0 | \vecx,m ]$ is equivalent to thresholding~$R$.
 It holds that $\pr[H_0 | \vecx,m ] \propto f( m| \vecx, H_0)$,
where the proportionality factor does not depend on~$\ell$.
We have
$f( m| \vecx, H_0) = f_{Y|X}(\vecm_\ell |\vecx)\prod_{j\neq \ell} f_Y(\vecm_j)$
$=\frac{f_{Y|X}(\vecm_\ell |\vecx)}{f_Y(\vecm_\ell)} \prod_j f_Y(\vecm_j)$
$\propto \frac{f_{Y|X}(\vecm_\ell |\vecx)}{f_Y(\vecm_\ell)}$ 
$\propto\exp(\frac{\vecm_\ell^2}{2\qs_Y^2}  - \frac{(\vecm_\ell-\vecx \sqrt T)^2}{2\qs_{Y|X}^2} ) $
$\propto \exp[  -\frac{ T\qs_X^2 }{2\qs_Y^2 \qs_{Y|X}^2} (\vecm_\ell - \vecx\sqrt T \frac{\qs_Y^2}{T \qs_X^2})^2 ]$
$\propto\exp[  J(\vecx, \vecm_\ell) \sqrt{n\qe}\sqrt{\frac{\vecx^2}{n\qs_X^2}(1+\qe)+\qe/2}]$,
again with proportionality factors that do not depend on~$\ell$.
The function $J$ is the score function defined in (\ref{defJ}).
We conclude that thresholding the score $J(\vecx,\vecm_\ell)$ is equivalent to thresholding the likelihood ratio~${\rm LR}$.

%=============================================================

%========================================================================

\bibliographystyle{unsrt}

\bibliography{randomcoding}
%========================================================================

\end{document}